\documentclass[aps,prl,twocolumn,floats,amsmath,preprintnumbers,nofootinbib,superscriptaddress,amssymb,10pt]{revtex4-2}
\usepackage{graphicx}
\usepackage{dcolumn}
\usepackage{bm}
\usepackage{hyperref}
\usepackage{cleveref}
\usepackage{xcolor}
\usepackage{physics}
\usepackage{multirow}
\usepackage{orcidlink}
\usepackage{amsfonts}
\usepackage{subcaption}
\usepackage[normalem]{ulem}

\DeclareMathOperator{\Orth}{O}

\newcommand{\s}{\mathbf{s}}
\newcommand{\nstep}{n_{\mathrm{step}}}
\newcommand{\ndof}{n_{\mathrm{dof}}}

\newcommand{\ESS}{\mathrm{ESS}}
\newcommand{\Var}{\mathrm{Var}}

\newcommand{\fex}{f_{\text{ex}}}
\newcommand{\fbulk}{f_{\text{bulk}}}
\newcommand{\Jfix}{J_{\text{fixed}}}
\newcommand{\Jonoff}{J_{\text{on}\rightarrow\text{off}}}
\newcommand{\Joffon}{J_{\text{off}\rightarrow\text{on}}}
\newcommand{\init}{\mbox{\tiny{i}}}
\newcommand{\fin}{\mbox{\tiny{f}}}

\begin{document}

\setcounter{secnumdepth}{3}

\setlength\abovedisplayskip{10pt}
\setlength\belowdisplayskip{10pt}

\setlength{\parskip}{4pt}
\setlength{\parindent}{0pt}

\preprint{HIP-2025-15/TH}


\title{Casimir effect in critical $\Orth(N)$ models from non-equilibrium Monte~Carlo simulations}

\author{Andrea Bulgarelli\orcidlink{0009-0002-2917-6125}}\email{andrea.bulgarelli@unito.it}
\affiliation{Department of Physics, University of Turin and INFN, Turin division, Via Pietro Giuria 1, I-10125 Turin, Italy}
\affiliation{Transdisciplinary Research Area ``Building Blocks of Matter and Fundamental Interactions'' (TRA Matter) and Helmholtz Institute for Radiation and Nuclear Physics (HISKP), University of Bonn, Nussallee 14-16, 53115 Bonn, Germany}

\author{Michele Caselle\orcidlink{0000-0001-5488-142X}}
\affiliation{Department of Physics, University of Turin and INFN, Turin division, Via Pietro Giuria 1, I-10125 Turin, Italy}

\author{Alessandro Nada\orcidlink{0000-0002-1766-5186}}
\affiliation{Department of Physics, University of Turin and INFN, Turin division, Via Pietro Giuria 1, I-10125 Turin, Italy}

\author{Marco Panero\orcidlink{0000-0001-9477-3749}}
\affiliation{Department of Physics, University of Turin and INFN, Turin division, Via Pietro Giuria 1, I-10125 Turin, Italy}
\affiliation{Department of Physics and Helsinki Institute of Physics, PL 64, FIN-00014 University of Helsinki, Finland}

\date{\today}

\begin{abstract}
$\Orth(N)$ vector models in three dimensions, when defined in a geometry with a compact direction and tuned to criticality, exhibit long-range fluctuations which induce a Casimir effect. The strength of the resulting interaction is encoded in the excess free-energy density, which depends on a universal coefficient: the Casimir amplitude. We present a high-precision numerical calculation of the latter, by means of a novel non-equilibrium Monte~Carlo algorithm, and compare our findings with results obtained from large-$N$ expansions and from the conformal bootstrap.
\end{abstract}

\maketitle

\section{Introduction}

Statistical systems of $N$-component unit vector spins $\s_i$, defined on the sites $i$ of a regular lattice and coupled via nearest-neighbor interactions through the globally $\Orth(N)$ symmetric Hamiltonian
\begin{align}
\label{Hamiltonian}
H = - J \sum_{\langle i, j \rangle} \s_i \cdot \s_j 
\end{align}
encompass many well-studied models~\cite{Stanley:1968ef, Stanley:1968gx}. For $N=0$, eq.~\eqref{Hamiltonian} describes the self-avoiding random-walk model, while for $N=1$ it reduces to the Hamiltonian of the Ising model (with applications for continuous transitions in physical systems as diverse as uniaxial antiferromagnets, liquid-gas transitions in fluids and binary fluid mixtures); for $N=2$, it yields the XY model (relevant for helium), while for $N=3$ it is commonly referred to as the Heisenberg model of ferromagnetism. The $N=4$ case is of relevance in the context of the Standard Model of elementary particle physics, where it can be associated with the universality class of the finite-temperature chiral phase transition of quantum chromodynamics with two light-quark flavors~\cite{Pisarski:1983ms}, or with a toy model for the Higgs sector in the Standard Model of elementary particle physics, whereas for $N=5$ it has been argued that it may describe high-temperature superconductivity~\cite{Hu:2000bat}, although this has later been refuted~\cite{Aharony:2002zz, Calabrese:2002bq, Hasenbusch:2005nx}. Finally, in the $N\to\infty$ limit eq.~\eqref{Hamiltonian} reduces to the spherical model: in this limit the theory has higher spin symmetry, and it has been conjectured~\cite{Klebanov:2002ja} that its singlet sector has a holographic dual equivalent to a Vasiliev theory~\cite{Vasiliev:1995dn}. In two dimensions, analytical solutions for some of these models have been known for many decades~\cite{Onsager:1943jn, Berezinsky:1970fr, Kosterlitz:1973xp, Kosterlitz:1974nba, Jose:1977gm, Nienhuis:1982fx, Nienhuis:1984wm}. More recently, significant progress has been achieved also in three dimensions: in particular, the critical $\Orth(N)$ conformal field theory (CFT) has been studied by means of $\epsilon$ expansions~\cite{Wilson:1971dc, Shalaby:2020xvv}, lattice high-temperature expansions~\cite{Butera:1997ak, Guida:1998bx, Pelissetto:2000ek, Campostrini:2002cf, Campostrini:2006ms, Butera:2007jd}, large-$N$ expansions~\cite{Vasiliev:1981dg, Vasiliev:1982dc, Petkou:1995vu, Derkachov:1997ch, Moshe:2003xn, Manashov:2017xtt, Alday:2019clp} and the conformal bootstrap~\cite{Kos:2014bka, Kos:2015mba, Kos:2016ysd}. Recent works have also focused on defects in this class of models~\cite{Cuomo:2021kfm,Cuomo:2022xgw,Bianchi:2022sbz,Trepanier:2023tvb,Raviv-Moshe:2023yvq,Giombi:2023dqs,Harribey:2023xyv,Bianchi:2023gkk}.

When a three-dimensional system described by the Hamiltonian~\eqref{Hamiltonian}, tuned to criticality, is defined in a finite spatial volume, the critical fluctuations induce long-range interactions between the boundaries: this phenomenon, akin to the Casimir effect (but of thermal, rather than quantum, origin~\cite{Schwinger:1977pa, FisherDeGennes1978}), is the main focus of the present work. 

Remarkably, such behavior is insensitive to the specific details of the system, in practice the microscopic degrees of freedom. In this sense, the Casimir effect displays a universal behavior, which only depends on the dimensionality of the system under study, the symmetries, and the geometrical details, such as the presence of boundaries.

In what follows, we consider a system defined on a cubic lattice of sizes $L\times L \times l$ (with $L\gg l$) with periodic boundary conditions along the three main axes; when the model is tuned to its bulk critical point, the scaling behavior is governed by a thermal conformal field theory, with $l$ playing the role of the inverse temperature of the underlying continuum field theory. As a consequence of the separation between the $l$ and $L$ scales, the free energy can be written as
\begin{align}
    \label{eq:free_energy}
    F(L,l) \equiv L^2 l f = L^2\left[\beta^{-1}\fex + l \fbulk(L)\right],
\end{align}
where $\fbulk$ is the free-energy density $f$ for an isotropic lattice, i.e. $l=L$, when the limit $L\rightarrow\infty$ is taken, to suppress further finite-size effects. The excess free-energy density $\fex$ quantifies the deviation from the isotropic limit, and encodes information on the thermal behavior of the CFT. In the scaling limit, it takes the simple form
\begin{align}
\label{eq:delta}
    \fex(l) = \Delta l^{-2},
\end{align}
where $\Delta$ is the \textit{critical Casimir amplitude}. Such quantity is universal, namely it is the same within a given universality class. The fact that critical exponents and amplitudes (or ratios of thereof) are insensitive to the microscopic details of a theory is a central concept in the theory of critical phenomena.

On a more formal level, $\Delta$ is part of the CFT data of the theory, which completely specify the universality class and quantitatively determine the long-range physics. In particular, $\Delta$ is the amplitude appearing in the definition of the one-point function of the stress-energy tensor in a thermal background~\cite{Luo:2022tqy, Sun:2024qhv, Barrat:2024fwq}. The large-$N$ limit of the critical Casimir amplitude is~\cite{Dantchev:1998etd}
\begin{align}
\label{eq:delta_largeN}
    \lim_{N\rightarrow\infty}\frac{\Delta}{N} = -\frac{2\zeta(3)}{5\pi} = -0.153050638412\dots,
\end{align}
where $\zeta$ denotes the Riemann zeta function; however its value for generic $N$ is not known analytically. As will be discussed in detail below, previous numerical calculations showed that, while the value of $\Delta/N$ in the Ising model is close to the prediction in eq.~\eqref{eq:delta_largeN}, sizeable deviations appear for $N>1$. In fact, the leading correction in $1/N$ was recently computed in ref.~\cite{Diatlyk:2023msc} and found to be rather large. In this work, we report the results of a high-precision numerical study of the Casimir amplitude for $N=1$, $2$, $3$, $4$, and $6$, obtained by means of an algorithm combining non-equilibrium Monte~Carlo (NEMC) simulations with a novel simulation technique to evaluate (derivatives of) the free-energy density.

\section{Related work} 

Markov chain Monte Carlo (MCMC) simulations have been widely employed for the study of the Casimir effect in lattice spin models. The determination of $\Delta$ in thin-film geometries with periodic boundary conditions has been carried out in refs.~\cite{PhysRevE.53.4414, Krech_1997, Dantchev:2004sy, Vasilyev_2007, Hucht:2010su, Vasilyev_2009, Hasenbusch:2009} for the Ising model and the $\Orth(2)$ spin model, leveraging standard Monte Carlo methods to compute free energy differences. In particular, periodic boundary conditions enable the study of conformal field theories at finite temperature, and allow one to make precise comparison with large-$N$ calculations. It is worth mentioning that the choice of boundary conditions crucially determines the nature of the critical Casimir effect. From an experimental standpoint, the most relevant boundary conditions are Dirichlet and open ones, as they can be directly realized in laboratory settings. In this context, not only the critical Casimir amplitude but also the entire scaling function, governing both the critical and off-critical behavior of the Casimir effect, becomes of primary interest. These types of boundary conditions have also been extensively investigated with MCMC simulations~\cite{Hucht:2007, Vasilyev_2009, Hasenbusch_2009, Hasenbusch:2009fb, Hasenbusch_2010, Parisen_Toldin_2010, Hasenbusch_2012, Diehl:2014bpa, Hasenbusch_2015}. Importantly, even though in the present work we only focus on periodic boundary conditions, in order to compare with analytical results, the applied methodology can be straightforwardly extended to arbitrary boundaries.

Recent developments have led to the first determinations of the critical Casimir amplitude using conformal bootstrap techniques~\cite{Iliesiu:2018zlz, Barrat:2024fwq}, yielding results that exhibit some tension with those obtained from Monte Carlo simulations. One of the purposes of the present work is precisely to quantify the significance of this discrepancy. For this reason, we devoted particular attention to the evaluation of both statistical and systematic uncertainties affecting our numerical results. Later, we will compare state-of-the-art MCMC results and recent bootstrap analyses with our findings.

Other studies of $\Delta$ in $\Orth(N)$ models, based on different methods, include those reported in refs.~\cite{PhysRevA.46.1886, Gruneberg:2007av, PhysRevE.53.2104, Dantchev:1998etd, Dantchev:2005pq, Jakubczyk:2012iza,Rancon:2016}; for reviews of the subject, see also refs.~\cite{Gambassi:2008beg, RevModPhys.90.045001, Dantchev:2022hvy, Gambassi:2024}.

In recent years, novel numerical methods were developed to efficiently compute differences of free energies on the lattice. A line of research builds on results from non-equilibrium statistical mechanics, in particular Jarzynski's~\cite{Jarzynski:1996oqb} and Crooks'~\cite{Crooks:1998} theorems, to construct a non-equilibrium Monte Carlo sampling scheme~\cite{Caselle:2016wsw}, enabling efficient calculations of free energy differences in a variety of contexts~\cite{Caselle:2018kap, Francesconi:2020fgi, Bulgarelli:2023ofi, Bonanno:2024udh, Bulgarelli:2024onj, Vadacchino:2024lob}. In particular it has been shown that NEMC methods are particularly well-suited to study statistical systems or quantum field theories with defects~\cite{Caselle:2016wsw}, or with non trivial geometric constraints and/or boundary conditions, as is the case of the problem we address in the present work. Furthermore, we note that non-equilibrium statistical mechanics theorems have also been recently employed not as purely algorithmic tools (as in this work), but to study the actual non-equilibrium quantum thermodynamics of lattice gauge theories~\cite{Davoudi:2024osg, Davoudi:2025tbi}.

Recently it was also shown that NEMC, which is essentially equivalent to annealing importance sampling~\cite{neal2001annealed}, shares the same theoretical background with certain machine-learning methods, namely normalizing flows~\cite{Albergo:2019eim, Nicoli:2020njz}, and can be combined with the latter to form stochastic normalizing flows~\cite{wu2020stochastic, Caselle:2022acb}, to systematically improve NEMC~\cite{Caselle:2024ent, Bulgarelli:2024yrz} (see also ref.~\cite{Albergo:2024trn} for a related approach). Crucially, non-equilibrium approaches have been shown to display a clear scaling with the number of degrees of freedom~\cite{Bonanno:2024udh, Bulgarelli:2024brv}.

\section{NEMC for Casimir amplitude computations}

\begin{figure*}[t]
    \centering
    \includegraphics[width=.8\linewidth]{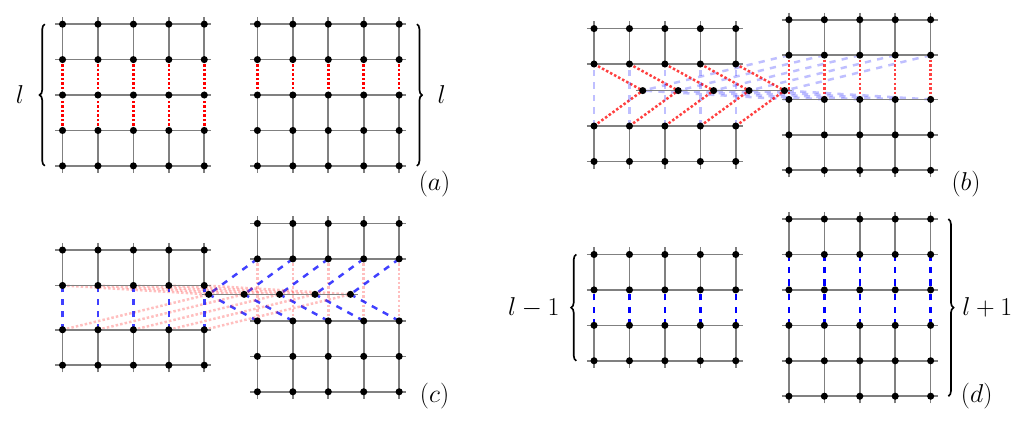}
    \caption{Protocol to compute second derivatives of free energies. Starting from two lattices of equal height $l$, panel (a), a slab is detached from the left-hand lattice, by removing a set of links (red dotted lines) with coupling $\Jonoff$. Simultaneously, a new set of links is introduced (blue dashed lines) with coupling $\Joffon$, embedding the slab in the right-hand lattice, panels (b) and (c). In the end, panel (d), the system consists of two lattices of heights $l-1$ and $l+1$ respectively.}
    \label{fig:slab_exchange_protocol}
\end{figure*}

The critical Casimir amplitude is encoded in the free energy of the model. Traditional Monte Carlo methods are typically not efficient in estimating the latter quantity, as the free energies (and differences of thereof) cannot be directly expressed as equilibrium averages. If one gives up the assumption of equilibrium MCMC, Jarzynski's equality~\cite{Jarzynski:1996oqb} can be used to directly estimate free energy differences from non-equilibrium averages.

In a non-equilibrium Monte Carlo simulation, a configuration sampled from an initial Boltzmann distribution $q_0=\exp(-\beta H_{\init})/Z_{\init}$ is driven towards a final distribution $p=\exp(-\beta H_{\fin})/Z_{\fin}$ using a protocol $c(n)$ that interpolates between $H_{\init}$ and $H_{\fin}$. In particular, this is implemented through a sequence of Monte Carlo updates each defined by the transition probability $P_{c(n)}$. The sequence of configurations reads:
\begin{equation}
\label{eq:NE_evol}
    \{\mathbf{s}\}_0  
    \stackrel{P_{c(1)}}{\longrightarrow} \{\mathbf{s}\}_1
    \stackrel{P_{c(2)}}{\longrightarrow} \dots
    \stackrel{P_{c(\nstep)}}{\longrightarrow} \{\mathbf{s}\}_{\nstep},
\end{equation}
where $\{\mathbf{s}\}_n$ denotes a configuration at step $n$. The difference of free energy between the final and initial distribution is computed using Jarzynski's equality~\cite{Jarzynski:1996oqb}
\begin{align}
\label{eq:jar}
    F_{\fin} - F_{\init} = -\frac{1}{\beta} \log\left\langle e^{-\beta W} \right\rangle_{\rm f},
\end{align}
where $\langle ... \rangle_{\rm f}$ is the average over an ensemble of the non-equilibrium evolutions of eq.~\eqref{eq:NE_evol} and $W$ the work done on the system
\begin{align}
    W = \sum_{n=0}^{\nstep-1} H_{c(n+1)}(\{\mathbf{s}_n\}) - H_{c(n)}(\{\mathbf{s}_n\}),
\end{align}
with $H_{c(\nstep)} \equiv H_{\fin}$ and $H_{c(0)} \equiv H_{\init}$. 
A relevant metric to quantify the efficiency of eq.~\eqref{eq:jar} in determining differences in free energy is the effective sample size (ESS), defined as
\begin{align}
\label{eq:ess}
    \ESS = \frac{\left\langle e^{-\beta W} \right\rangle^2}{\left\langle e^{-2 \beta W} \right\rangle} = \frac{1}{\left\langle e^{-2 \beta (W - \Delta F)} \right\rangle}.
\end{align}
The $\ESS$ is a different way to express the variance of the estimator of eq.~\eqref{eq:jar}, as $\Var(e^{-\beta W}) = (1/\ESS -1) \, e^{-2 \beta\Delta F}$, which immediately indicates $0 \leq \ESS \leq 1$. For a fixed statistics, i.e., a fixed number of non-equilibrium evolutions, the estimator of eq.~\eqref{eq:jar} depends on how well the distribution of $e^{-\beta W}$ has been sampled: being the average of an exponential, it could problematic if the tails are not populated enough, ref.~\cite{Jarzynski_2006} for a related discussion. If the $\ESS$ is large enough, it guarantees that the distribution is sampled well enough and that the resulting estimate of the free-energy difference is accurate also with limited statistics. Furthermore, controlling the $\ESS$ (or, equivalently, the relative variance) as a function of $\nstep$ provides a way of calibrating the algorithm: a detailed analysis of this aspect is presented in appendix~\ref{app:calibration}.

In ref.~\cite{Vasilyev_2009}, $\Delta$ was computed with a Monte Carlo algorithm estimating the first derivative of~\eqref{eq:free_energy} with respect to $l$; since the derivative does not allow one to isolate $\fex$, an iterative procedure was employed to remove $\fbulk$. In this work, we propose a method based on the calculation of the \textit{second} derivative, which automatically gets rid of the undesired bulk term.

Traditional calculations of second derivatives on the lattice require the subtraction of two first derivatives, and this can lead to larger errors, as they are not generally expressed as primary observables. Here we develop, for the first time, an algorithm for a direct calculation of second derivatives from ensemble averages. We start noticing that the lattice discretization of the second derivative of eq.~\eqref{eq:free_energy} with respect to $l$ reads
\begin{align}
    \eval{\pdv[2]{F}{l}}_{\text{lattice}} = F(l+1) + F(l-1) - 2F(l).
    \label{eq:second_derivative_lattice}
\end{align}
This quantity can be connected with the first derivative of eq.~\eqref{eq:delta}, thus allowing a direct determination of $\Delta$.

We set $F_{\init}=2F(l)$, representing two independent lattices with height $l$, and $F_{\fin}=F(l+1)+F(l-1)$, corresponding to a final distribution with lattices of heights $l+1$ and $l-1$ respectively. We introduce a \textit{slab-exchange protocol} connecting the two probability distributions as shown in fig.~\ref{fig:slab_exchange_protocol}. The couplings of the two lattices are partitioned into three sets: $\Jfix$ (solid gray lines), $\Jonoff$ (dotted red lines) and $\Joffon$ (dashed blue lines). The $\Jfix$ couplings remain fixed to $J$ throughout the protocol. The $\Jonoff$ couplings are linearly interpolated from $J$ to $0$, detaching a slab from the first lattice. Simultaneously, $\Joffon$ couplings are linearly interpolated from $0$ to $J$, thereby inserting the slab into the second lattice. Formally, the protocol is
\begin{align}
    c(n) = \begin{cases}
        \Jfix(n) &= J, \\
        \Jonoff(n) &= J\left(1-n/\nstep\right),\\
        \Joffon(n) &= Jn/\nstep.
    \end{cases}
\end{align}

Using Jarzynski's equality, the quantity in eq.~\eqref{eq:second_derivative_lattice} is computed as an average over non-equilibrium processes quenching the geometry of the lattice. This has the advantage of expressing the second derivative of the free energy as a primary observable (i.e. a quantity directly estimated from field configurations, in contrast to a secondary observable, which is constructed from multiple and possibly uncorrelated observables); if it was expressed as the difference of two first derivatives, the uncertainty on this quantity would be larger, as it would result from the combination of two uncorrelated primary observables. We also note that the first derivative of $F$ can be readily computed with non-equilibrium Monte Carlo evolutions as well, and the procedure followed in ref.~\cite{Vasilyev_2009} closely replicated. 

We would like to emphasize that Jarzynski's equality provides a natural and efficient framework to compute free-energy differences in which every source of error is fully understood using non-equilibrium statistical mechanics. First, we stress that eq.~\eqref{eq:jar} is an unbiased estimator of $\Delta F$ for any protocol, e.g., for any value of $\nstep$. As mentioned before though, results can be considered accurate only if the exponential average is sampled well enough with limited statistics. This issue can be first monitored within the algorithm itself by computing the $\ESS$ and eventually solved by selecting a large enough value of $\nstep$. Crucially, the same metric can also be used to optimize directly the non-equilibrium algorithm itself to minimize the relative error on $\Delta F$.

Furthermore, it is useful to compare this approach with the integral method used e.g. in refs.~\cite{Vasilyev_2009, Hasenbusch_2009, Hasenbusch_2010}. Interestingly, they essentially coincide in the infinite intermediate steps limit: the evolutions would be at equilibrium and the integral giving the free energy difference would be calculated exactly with no systematic errors. Away from this special case though, the integral method is very different, as the free energy is not obtained directly, but requires the integration of statistically independent expectation values. This has several drawbacks: first, the systematic error induced by the numerical integration has to be controlled in each case and depends strongly on the method used for the integration; second, each integrand comes from an independent simulation that has to be thermalized (a process not needed in NEMC); third, it is not immediate to define the most efficient setup for the numerical integration itself. All these issues are addressed directly within the NEMC framework, which provides a powerful, self-contained toolbox that allows to easily tackle more challenging computations such as the one of eq.~\eqref{eq:second_derivative_lattice}.

As a final comment of this section, we emphasize that the slab-exchange algorithm can be generalized to different choices of boundary conditions. If the boundaries break translational invariance in the short direction, which is the case for, e.g., Dirichlet or open boundary conditions, one has to specify where the slab is removed from the first lattice and where it is inserted in the second one. While from a theoretical perspective the choice of the slab is immaterial, as a consequence of the Jarzynski's theorem, different protocols can lead to varying levels of efficiency.

\section{Numerical results}

We aim to compute numerically the critical Casimir amplitude $\Delta$ for several $\Orth(N)$ models, up to $N=6$. 
For every $N$, we tuned the system to the bulk critical point $\beta_c$ (see appendix~\ref{app:details} for the values we used) and computed the second derivative of the excess of free energy density $\partial^2 \fex/\partial l^2 = \frac{\beta}{L^2} \partial^2 F/\partial l^2 $ with the slab-exchange algorithm. We refer to appendix~\ref{app:calibration} for a detailed discussion on the optimization of the $\nstep$ parameter in the non-equilibrium evolutions to compute the second derivative of $F$.
The Monte Carlo updating algorithm used in all numerical simulations is an embedding cluster algorithm~\cite{Wolff:1988uh, Wolff:1988kw, Wolff:1989hv, Hasenbusch:1989kx} highly parallelized on GPUs with CUDA~\cite{Komura_2012,Komura_2014}. 

To determine $\Delta$ a careful analysis of the finite-size effects is required. First, we performed an extrapolation $L\rightarrow\infty$. The scaling corrections to the free energy when $L\gg l$ are expected to be exponential in $L$; therefore, we used the \textit{Ansatz} 
\begin{align}
    \frac{\partial^2 \fex}{\partial l^2}(L;A,k,m) = A\exp(-mL) + k,
    \label{eq:large_L_fit_function}
\end{align}
where $A$, $k$ and $m$ are fitting coefficients, the first two depending on $l$, while the latter is a global parameter. For every $l$, we quote as $\partial^2 \fex/\partial l^2$ the result we obtained from the simulation with the largest $L$, with a systematic error given by the difference with the asymptote $k$ obtained from the fitting procedure. The systematic error is summed in quadrature with the statistical one.

\begin{figure}
    \centering
    \includegraphics[width=.8\linewidth]{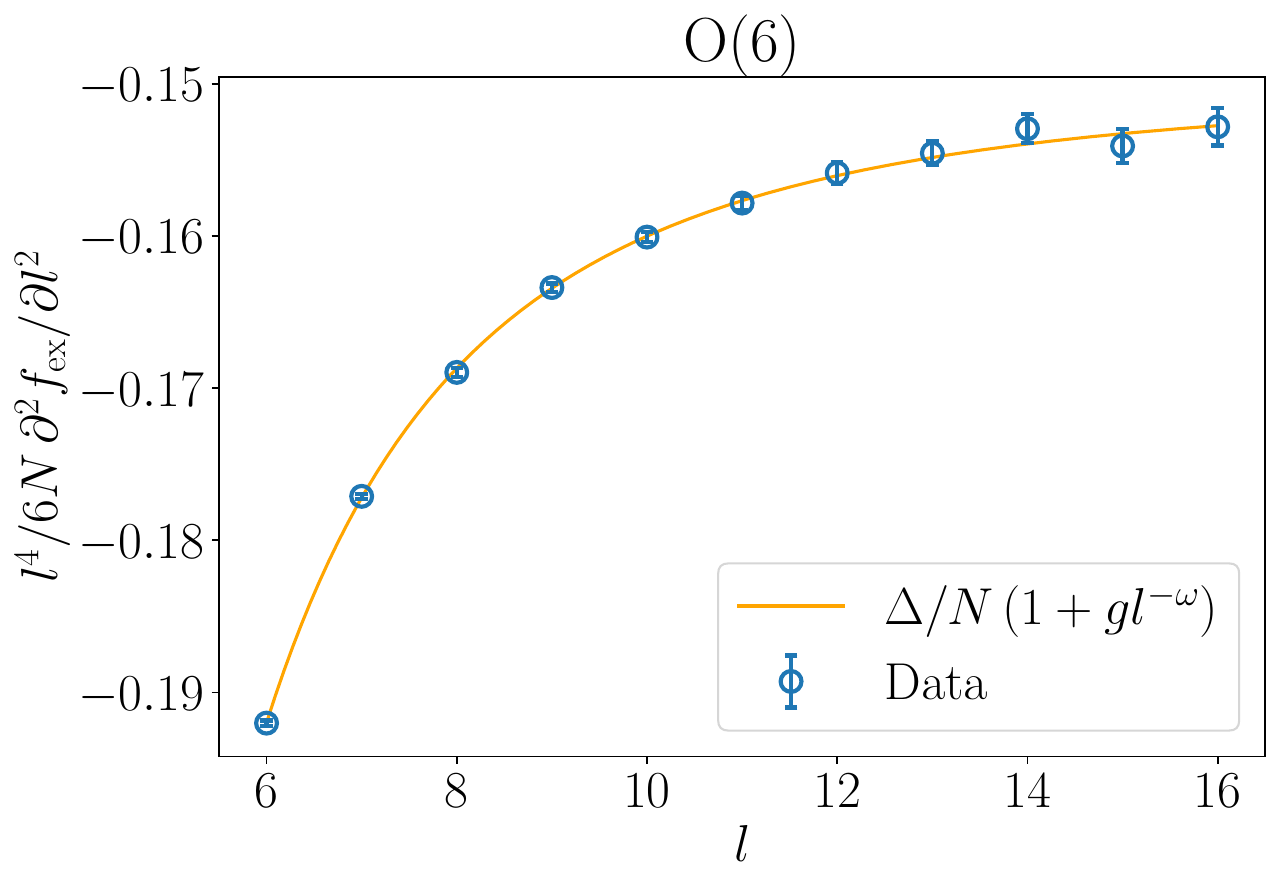}
    \caption{Results of $\partial^2\fex/\partial l^2$ (blue circles) for various values of the lattice size $l$ for $N=6$, appropriately normalized. The orange curve is the best fit result for the functional form of eq.~\eqref{eq:fit_function_l}.}
    \label{fig:fit_l_O6}
\end{figure}

The value of $\Delta$ is extracted with a fit in $l$ of the values of $\partial^2 \fex/\partial l^2$ obtained in the previous step. Here we use the same fit function used in ref.~\cite{Vasilyev_2009}, motivated by a finite-size scaling analysis
\begin{align}
    \frac{\partial^2 \fex}{\partial l^2}(l;\Delta,g,\omega) = 6\Delta \, l^{-4}(1+g\, l^{-\omega}),
    \label{eq:fit_function_l}
\end{align}
where the leading term $l^{-4}$ is given by the second derivative of eq.~\eqref{eq:delta} and $g$ and $\omega$ control the leading finite-size corrections. It is important to emphasize that the functional form~\eqref{eq:fit_function_l} is not unique. Therefore, similarly to past determinations that are present in literature~\cite{Vasilyev_2009}, our result is biased by the specific choice of the fit function~\eqref{eq:fit_function_l}.
In fig.~\ref{fig:fit_l_O6} we show the result of this fit for the $N=6$ case. As shown in table~\ref{tab:fit_l}, the quality of the fit is very good for all values of $N$. It is worth commenting on the best-fit results for the scaling exponent $\omega$. For all the values of $N$, the exponents are compatible with each other, with the exception of $N=6$, which nevertheless remains consistent within two standard deviations. Currently, we are not aware of a clear theoretical justification for $\omega$ to take the same value across different $N$. Therefore, in our analysis, we treat $\omega$ for different $N$ as different parameters. We also notice that the values obtained in the present study are close, yet not compatible with the values of $\omega$ found in~\cite{Vasilyev_2009}, in particular $\omega_{\Orth(1)} = 2.664(27)$ and $\omega_{\Orth(2)}=2.59(4)$. We refer to appendices~\ref{app:bigL} and~\ref{app:smalll} for further details on the fitting procedures.

\begin{table*}[t]
    \centering
    \begin{tabular}{|c|c|c|c|c|c|}
    \hline
    & $\Orth(1)$ & $\Orth(2)$ & $\Orth(3)$ & $\Orth(4)$ & $\Orth(6)$ \\
    \hline\hline
    $\chi^2_{\mathrm{red}}[\#_{\mathrm{dof}}]$ & $1.18$ $[12]$ & $0.21$ $[8]$ & $1.33$ $[8]$ & $0.78$ $[8]$ & $0.46$ $[8]$\\
    \hline
    $g$ & $57(4)$ & $63(10)$ & $59(9)$ & $49(9)$ & $44(6)$\\
    \hline
    $\omega$ & $3.02(5)$ & $3.08(10)$ & $3.01(10)$ & $2.90(11)$ & $2.83(8)$\\
    \hline
    \end{tabular}
    \caption{Reduced $\chi^2$ and fit parameters for different extrapolations in $l$ using eq.~\eqref{eq:fit_function_l}.}
    \label{tab:fit_l}
\end{table*}

\begin{figure}
    \centering
    \includegraphics[width=\linewidth]{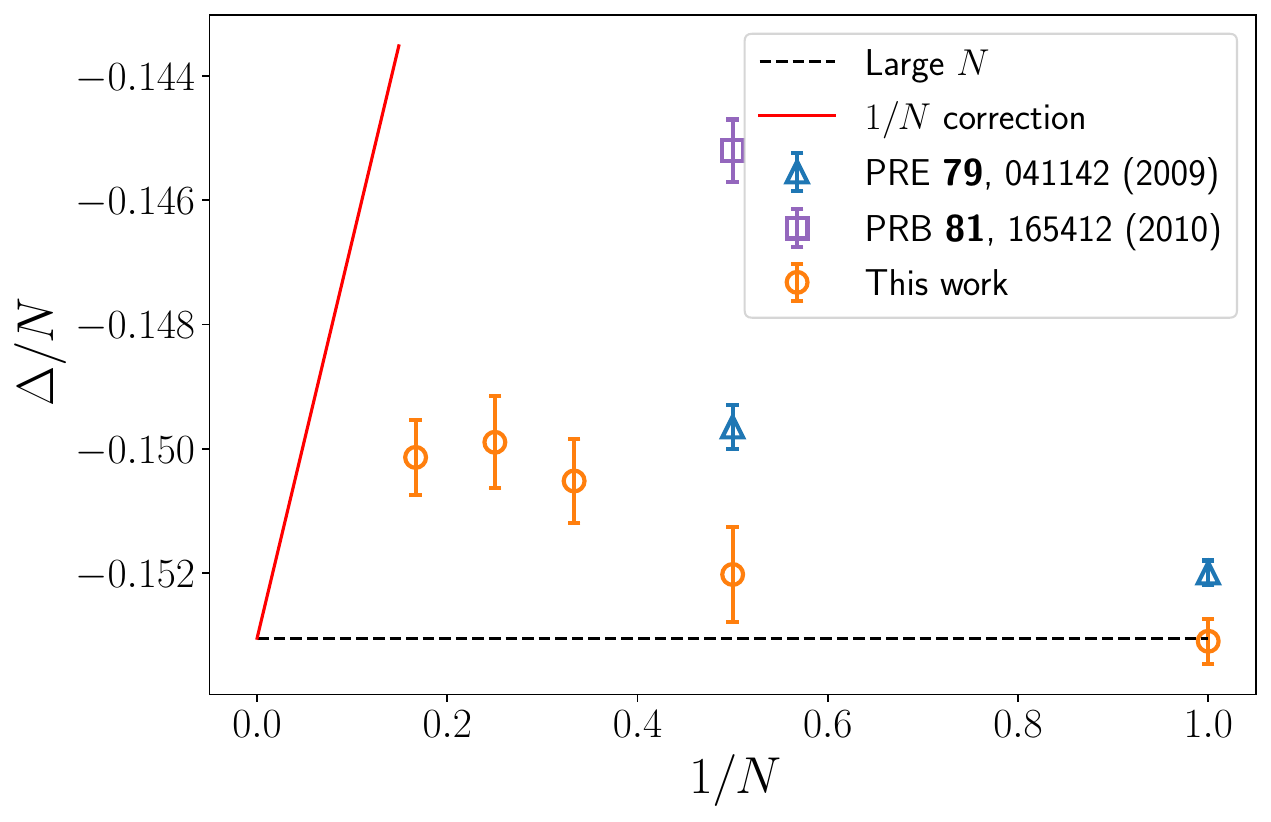}
    \caption{Numerical estimates of $\Delta/N$ (orange circles) compared with previous Monte Carlo results~\cite{Vasilyev_2009} (cyan triangles) and large-$N$ calculations~\cite{Dantchev:1998etd,Diatlyk:2023msc}.}
    \label{fig:amplitudes_different_N}
\end{figure}

The results for the amplitude $\Delta$ extracted from eq.~\eqref{eq:fit_function_l} for all values of $N$ taken into consideration are presented in fig.~\ref{fig:amplitudes_different_N} and listed in table~\ref{tab:results_Delta}. Our numerical determination can be compared with the behavior expected in the $N \to \infty$ limit from~\cite{Dantchev:1998etd}, see eq.~\eqref{eq:fit_function_l}: interestingly, the result for $N=1$ is remarkably close to the analytical prediction in the $N\to \infty$ limit, while the $N=2$ result clearly deviates from it. This was already observed in earlier numerical work for $N=1$ and $N=2$~\cite{Vasilyev_2009}, also shown in fig.~\ref{fig:amplitudes_different_N}: we point out that the difference between~\cite{Vasilyev_2009} and our determination is due to a different treatment of finite-size effects in eq.~\eqref{eq:large_L_fit_function}; we refer to appendix~\ref{app:bigL} for a detailed discussion of the matter. The more significant discrepancy with respect to other results from the Monte Carlo literature, in particular~\cite{Hasenbusch:2009}, might be due to the bias introduced by the particular functional form assumed for the finite-size correction~\eqref{eq:fit_function_l}. In light of this fact, and also the previous analysis of the critical exponent $\omega$, we expect a more precise theoretical understanding of such contributions, or the use of improved models where such corrections can be arbitrarily suppressed, to be beneficial for a more precise determination of the critical Casimir amplitude.

A clearer picture of the trend of the critical Casimir amplitude is provided by our numerical results for $N=3$, $4$, and $6$: on the one hand, the values of $\Delta/N$ are even larger than the result obtained for $N=2$, but on the other hand they are all largely compatible with each other within the statistical uncertainties. Although we are not yet able to determine in a conclusive manner whether and how the $N \to \infty$ limit is approached, we wish to point out that the first correction with respect to the large-$N$ limit, computed in ref.~\cite{Diatlyk:2023msc}, has a sign (and a size) compatible with our results. In particular, for $N=10$ it is already roughly consistent with our numerical results, as clearly displayed in fig.~\ref{fig:amplitudes_different_N}.

\begin{figure}
    \centering
    \includegraphics[width=\linewidth]{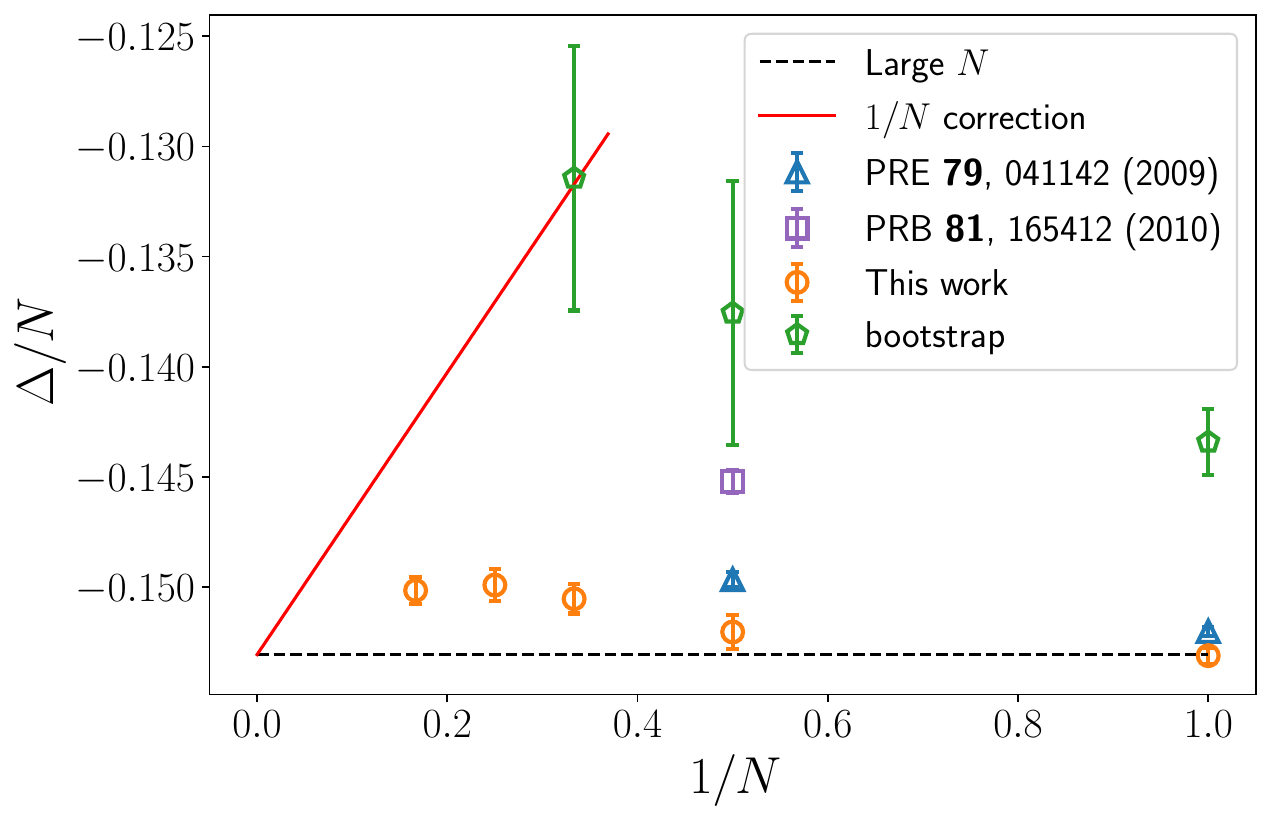}
    \caption{Comparison between different Monte Carlo determinations of $\Delta/N$ and the bootstrap result from~\cite{Barrat:2024fwq}.}
    \label{fig:amplitudes_bootstrap}
\end{figure}

It is important to observe that the results of thermal bootstrap~\cite{Barrat:2024fwq}, reported in table~\ref{tab:results_Delta}, show a discrepancy with the Monte Carlo determination, see fig.~\ref{fig:amplitudes_bootstrap}. Even though the qualitative trend is the same, with $\Delta/N$ increasing from $N=1$ to $N=3$, a systematic shift is manifest. We also stress that, in~\cite{Barrat:2024fwq}, the authors focused on the calculation of two-point functions coefficients, and the critical Casimir amplitude is then obtained indirectly through a conversion formula~\cite{privcomm}. A better understanding of such discrepancy might come from precision lattice estimates of the two point function of the models, enabling a more direct comparison with bootstrap results.

\begin{table}[t]
    \centering
    \begin{tabular}{|c|c|c|c|c|}
    \hline
    & This work & MCMC~\cite{Vasilyev_2009} & MCMC~\cite{Hasenbusch:2009} & bootstrap~\cite{Barrat:2024fwq}\\
    \hline\hline
    $\Orth(1)$ & $-0.1531(4)$ & $-0.1520(2)$ & $-$ & $-0.1425(15)$ \\
    \hline
    $\Orth(2)$ & $-0.1520(8)$ & $-0.1496(3)$ & $-0.1452(5)$ & $-0.137(6)$ \\
    \hline
    $\Orth(3)$ & $-0.1505(7)$ & $-$ & $-$ & $-0.131(6)$\\
    \hline
    $\Orth(4)$ & $-0.1499(7)$ & $-$ & $-$ & $-$\\
    \hline
    $\Orth(6)$ & $-0.1501(6)$ & $-$ & $-$ & $-$\\
    \hline
    \end{tabular}
    \caption{Values of $\Delta/N$ from various determinations.}
    \label{tab:results_Delta}
\end{table}

\section{Conclusions} 

In this contribution, we presented novel numerical results for the Casimir amplitude in critical $\Orth(N)$ models with $N=1$, $2$, $3$, $4$ and $6$, obtained by means of a new type of non-equilibrium Monte Carlo algorithm to compute second derivatives on the lattice. The results for $N=4$ and $6$ are new, and $N=3$ has never been determined before with Monte-Carlo-based methods.

NEMC, providing a well understood and scalable method to compute differences of free energies, combined with the slab-exchange algorithm that we introduced in this work, enables a precise treatment of the systematic effects in the calculation of the critical Casimir amplitude, leading to high-precision results.

Our findings, compared to analytical studies of ref.~\cite{Diatlyk:2023msc}, suggest that $N=6$ is not large enough to fully capture the large-$N$ behavior of $\Orth(N)$ models, and the behavior of $\Delta$ as a function of $N$ is non-monotonic. This can be contrasted with $\mathrm{SU}(N)$ gauge theories, where, for a wide variety of physical quantities, $N=3$ results are already close to the $N\rightarrow\infty$ limit~\cite{Panero:2009tv, Lucini:2012gg}. A better understanding of the dynamics of $\Orth(N)$ models requires an interplay between high-precision numerical results at larger values of $N$ and the determination of higher-order $1/N$ corrections.

\begin{acknowledgments}
    \section*{Acknowledgments}
    \noindent
    We are grateful to C.~Luo for insightful comments and correspondence and for pointing out reference~\cite{Diatlyk:2023msc}, and to A.~Miscioscia and E.~Pomoni for reading a preliminar version of the draft and for sharing their bootstrap results reported in table~\ref{tab:results_Delta}. We also acknowledge discussions with T.~Canneti, E.~Cellini, A.~Mariani, D.~Panfalone and L.~Verzichelli.  We acknowledge support from the SFT Scientific Initiative of INFN. M.~C. and A.~N. acknowledge support and A.~B. acknowledges partial support by the Simons Foundation grant 994300 (Simons Collaboration on Confinement and QCD Strings).  A.~N. acknowledges support from the European Union -- Next Generation EU, Mission 4 Component 1, CUP D53D23002970006, under the Italian PRIN ``Progetti di Ricerca di Rilevante Interesse Nazionale – Bando 2022'' prot. 2022ZTPK4E. The numerical simulations were run on machines of the Consorzio Interuniversitario per il Calcolo Automatico dell'Italia Nord Orientale (CINECA).
\end{acknowledgments}

\appendix


\section{Details on the simulation}
\label{app:details}

Simulations have been performed at the bulk critical point, $\beta = \beta_c$, whose values are listed in table~\ref{tab:beta_c}, along with the references where they were numerically determined.

\begin{table}[h]
    \centering
    \begin{tabular}{|c|c|c|}
    \hline
    & $\beta_c$ & Reference \\
    \hline\hline
    $\Orth(1)$ & $0.221654626$ & \cite{Ferrenberg_2018}\\
    \hline
    $\Orth(2)$ & $0.4541652$ & \cite{Deng_2005} \\
    \hline
    $\Orth(3)$ & $0.693002$ & \cite{Deng_2005} \\
    \hline
    $\Orth(4)$ & $0.93600$ & \cite{Butera:1998rk}\\
    \hline
    $\Orth(6)$ & $1.42865$ & \cite{Holtmann:2003he} \\
    \hline
    \end{tabular}
    \caption{Values of $\beta_c$ used in the simulations in this work.}
    \label{tab:beta_c}
\end{table}

In all the cases, the error is on the last significant digit. We checked in our simulations that the systematic effect, induced by the uncertainty in the determination of the value of $\beta_c$, is smaller than the precision of our results.

The data analysis has been done using pyerrors~\cite{Joswig:2022qfe}.

\section{Calibration of the algorithm}
\label{app:calibration}

\begin{figure*}[ht]
    \centering
    \begin{subfigure}{.32\linewidth}
    \includegraphics[width=1\linewidth]{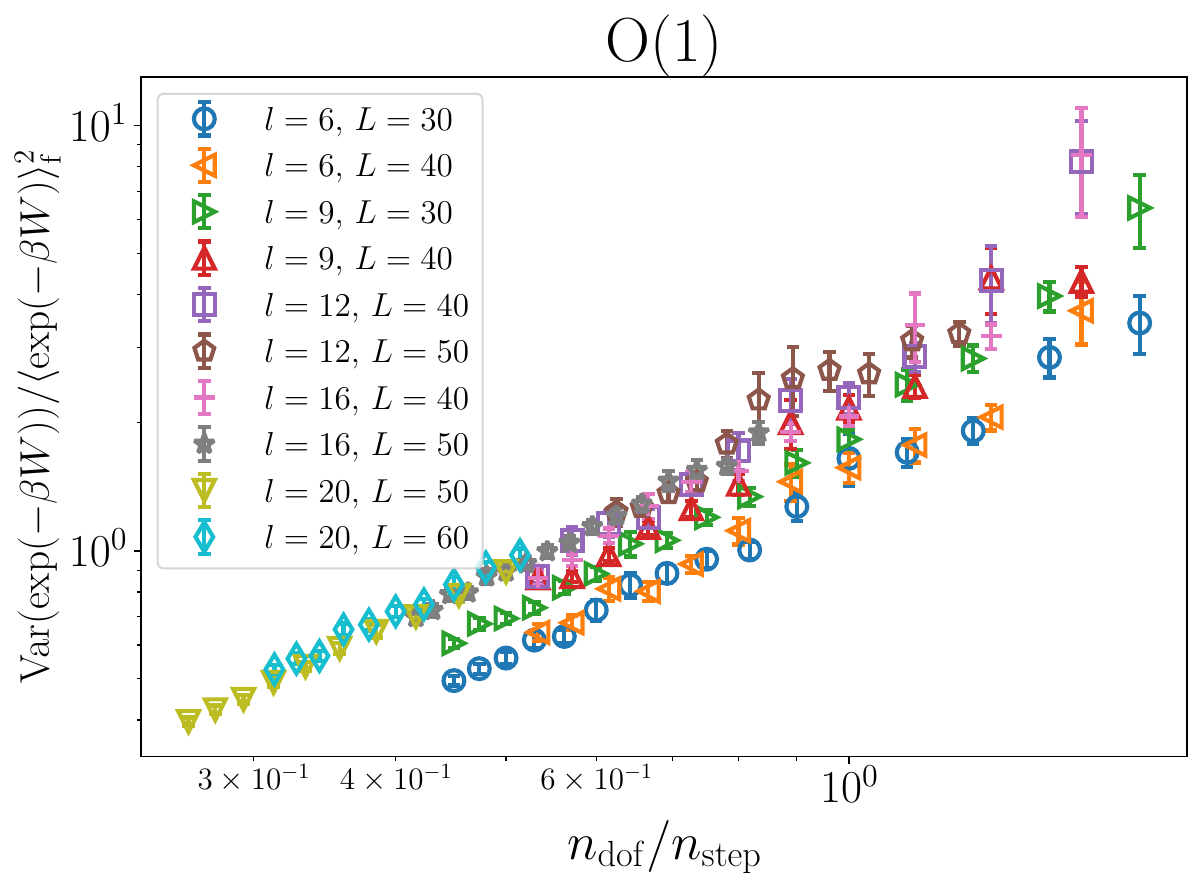}
    \end{subfigure}
    \begin{subfigure}{.32\linewidth}
    \includegraphics[width=1\linewidth]{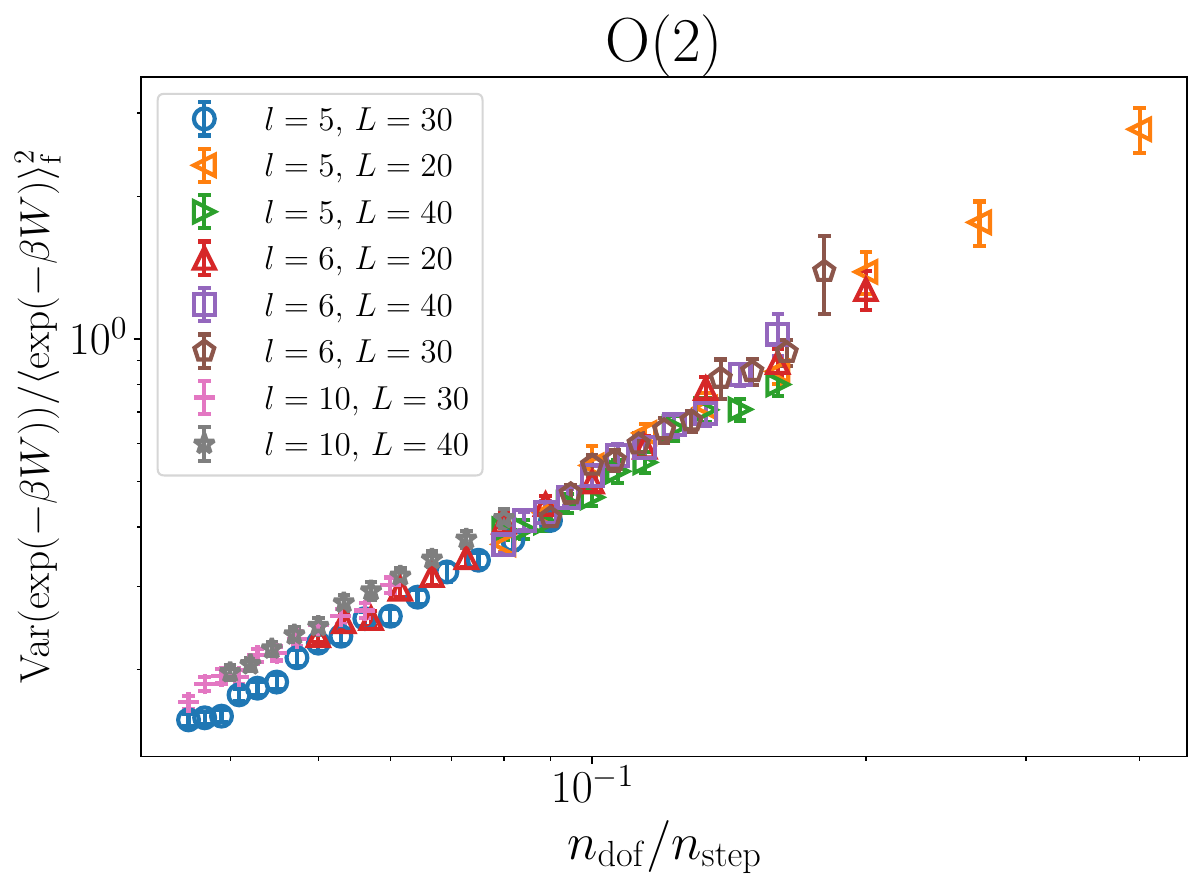}
    \end{subfigure}
    \begin{subfigure}{.32\linewidth}
    \includegraphics[width=1\linewidth]{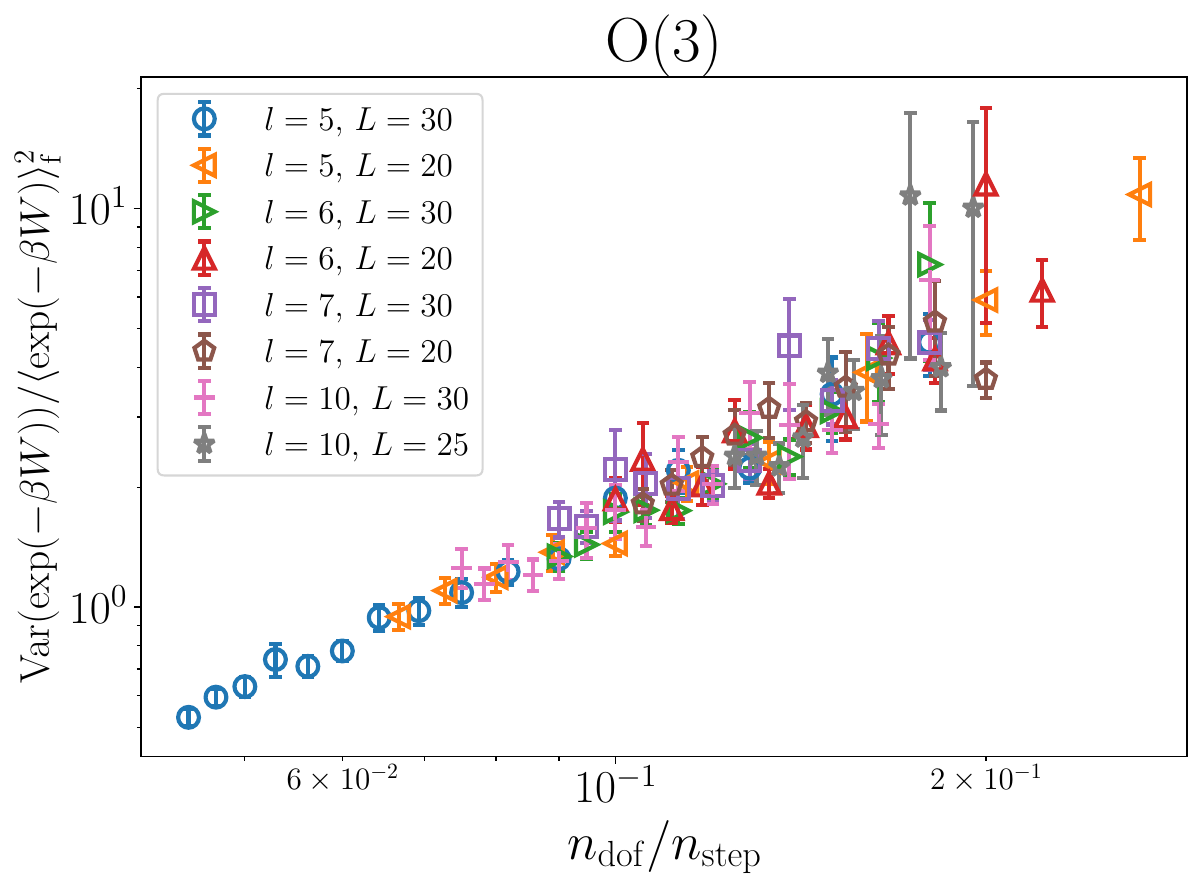}
    \end{subfigure}
    \begin{subfigure}{.32\linewidth}
    \includegraphics[width=1\linewidth]{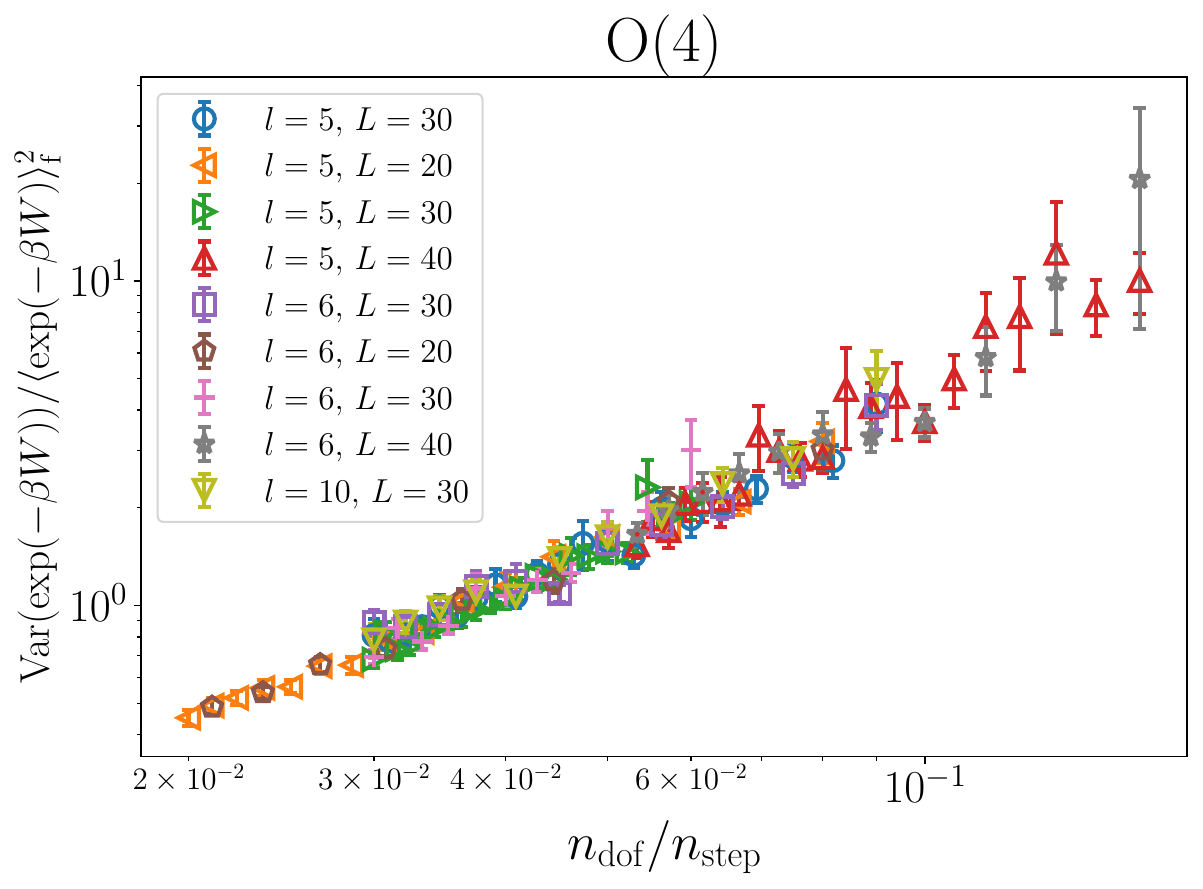}
    \end{subfigure}
    \begin{subfigure}{.32\linewidth}
    \includegraphics[width=1\linewidth]{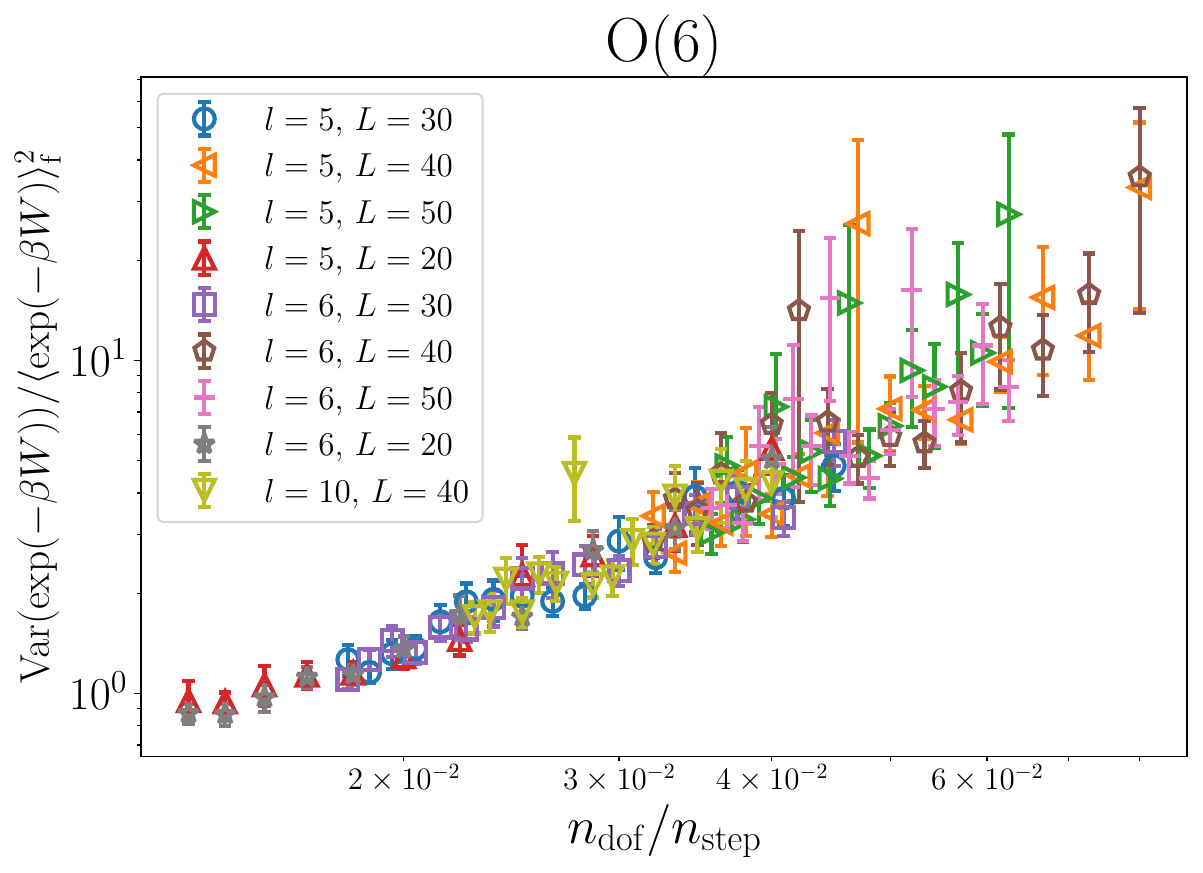}
    \end{subfigure}
    \caption{Calibration of the algorithm at the critical point $\beta=\beta_c$. The relative variance for different $\Orth(N)$ models is plotted against $\ndof/\nstep$. A data collapse is manifest for all the values of $N$ except for $N=1$.}
    \label{fig:calibration_collapse}
\end{figure*}

NEMC has been shown to exhibit a linear scaling with the number of degrees of freedom involved in the non-equilibrium evolutions~\cite{Bonanno:2024udh, Bulgarelli:2024brv}, here denoted as $\ndof$. In the present work, $\ndof$ is equal to the number of links undergoing the evolution, specifically the dotted blue and dashed red links of fig.~\ref{fig:slab_exchange_protocol}, and is proportional to the number of sites on a single slab, i.e., $L^2$. More precisely, the $\ESS$~\eqref{eq:ess}, i.e., the metric that we use to determine the sampling quality of a given protocol, is expected to be a function of the ratio $\ndof/\nstep$ only. If the system is close enough to the equilibrium, it is possible to make some assumptions on the specific functional dependence.

First, it is important to mention that the $\ESS$ is usually defined not for free energy calculations, but when sampling a given observable $\mathcal{O}$ according to the final distribution (in this case, $e^{-\beta H_{\mathrm{f}}}$). In this other context it represents an approximation for the ratio between the theoretical variance of $\mathcal{O}$ and the variance of the non-equilibrium estimator of $\mathcal{O}$: ref.~\cite{Bonanno:2024udh} contains a detailed analysis of this quantity in the context of NEMC.
However, for the scope of this work we can simply focus on the (exact) relation between the (relative) variance of $\exp(-\beta W)$ and the $\ESS$ that we discussed in the main text, namely
\begin{align}
    \frac{\Var(e^{-\beta W})}{\langle e^{-\beta W} \rangle_{\rm f}^2} = \frac{1}{\ESS}-1.
\end{align}
Intuitively, increasing $\nstep$ reduces the variance of $\exp(-\beta W)$, as the parameters are modified more slowly in the protocol and the system is closer to equilibrium throughout the evolution. Conversely, when $\ndof$ grows, larger fluctuations in the work distribution are expected. This motivates us to parametrize the relative variance as
\begin{align}
    \frac{\Var(e^{-\beta W})}{\langle e^{-\beta W} \rangle_{\rm f}^2} = \sum_{k = 1}^{k_{\rm max}} v_k \left(\frac{\ndof}{\nstep}\right)^k,
    \label{eq:calibration_equation}
\end{align}
where $k_{\rm max}$ is chosen so that the previous \textit{Ansatz} provides a good approximation of the numerical behavior of the $\ESS$. Empirically we found out that, at least for the purposed of this work, $k_{\rm max}=2$ is enough to approximate our data. Notice that in principle the coefficients $v_k$ are not constants, rather they may depend on some parameters of the theory, in our case $N$, $\beta$, $L$ and $l$. Remarkably, being able to fit the coefficients $v_k$ gives complete control over the algorithm. Indeed, given $\ndof$, one can use~\eqref{eq:calibration_equation} to tune $\nstep$ to reach a given $\ESS$.

Fig.~\ref{fig:calibration_collapse} shows the relative variance of $\exp(-\beta W)$ across the various models studied, plotted against $\nstep/\ndof$, for different values of $l$ and $L$. For $N\geq 2$, the data display a remarkable collapse for all the dataset we considered, allowing for a global calibration of the algorithm which is independent of the specific value of $l$. The Ising model ($N=1$) does not exhibit the same feature: although a clear data collapse is observed at fixed $l$, the resulting coefficients $v_k$ exhibit an explicit dependence on $l$. This still enables a calibration for every $l$.

\section{$L\rightarrow\infty$ extrapolation}
\label{app:bigL}

\begin{figure*}[ht]
    \centering
    \begin{subfigure}{.32\linewidth}
    \includegraphics[width=1\linewidth]{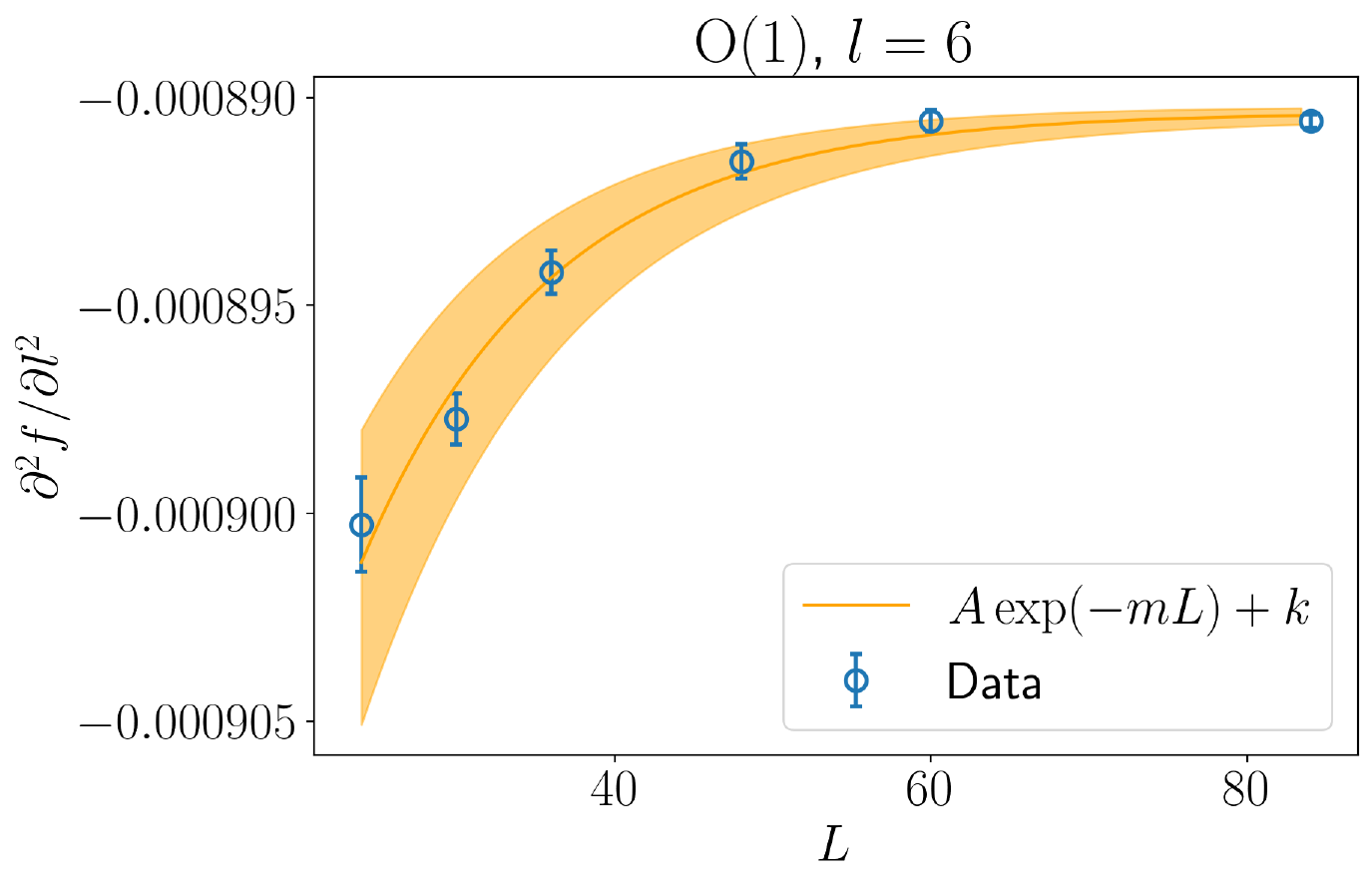}
    \end{subfigure}
    \begin{subfigure}{.32\linewidth}
    \includegraphics[width=1\linewidth]{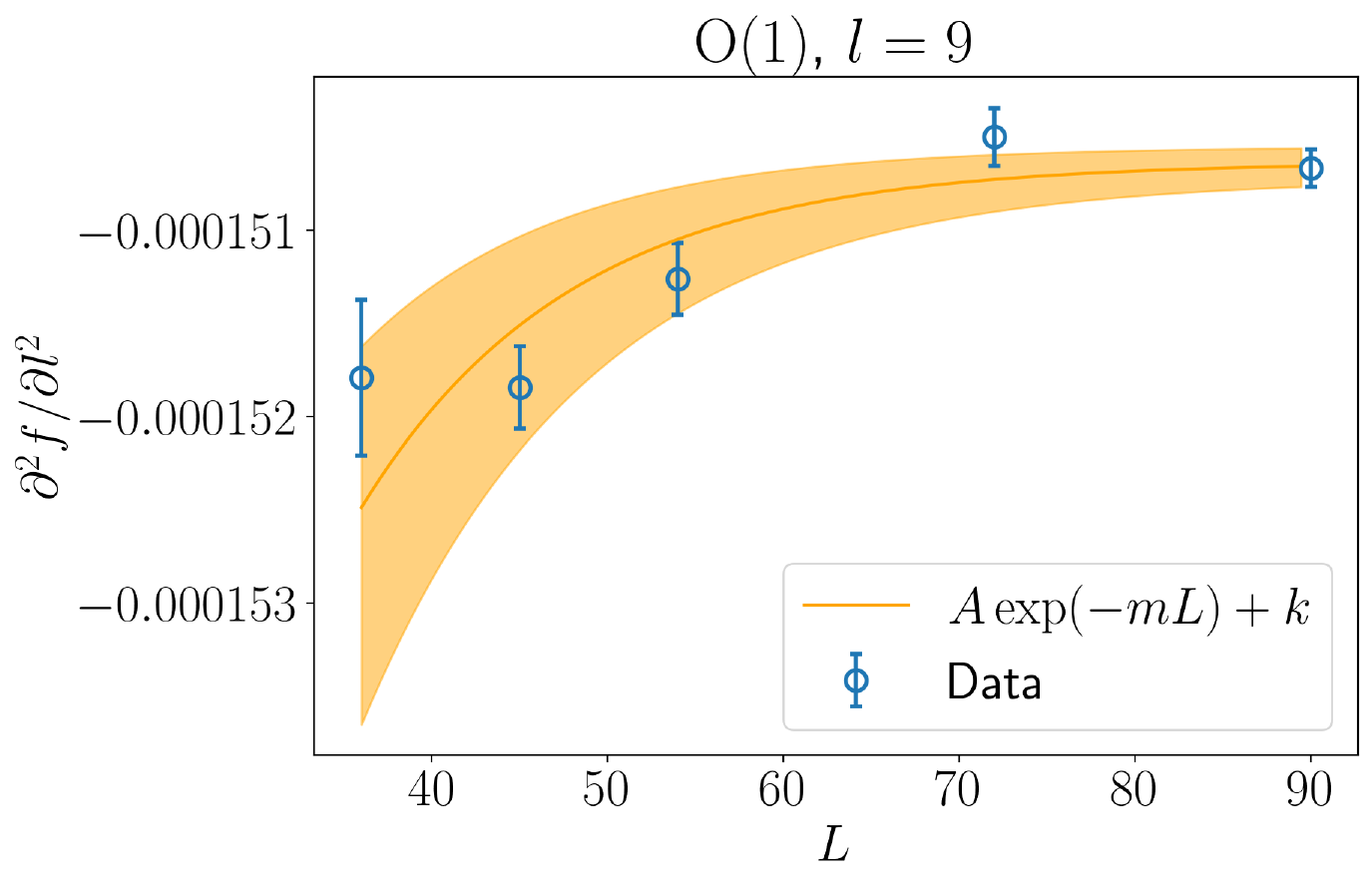}
    \end{subfigure}
    \begin{subfigure}{.32\linewidth}
    \includegraphics[width=1\linewidth]{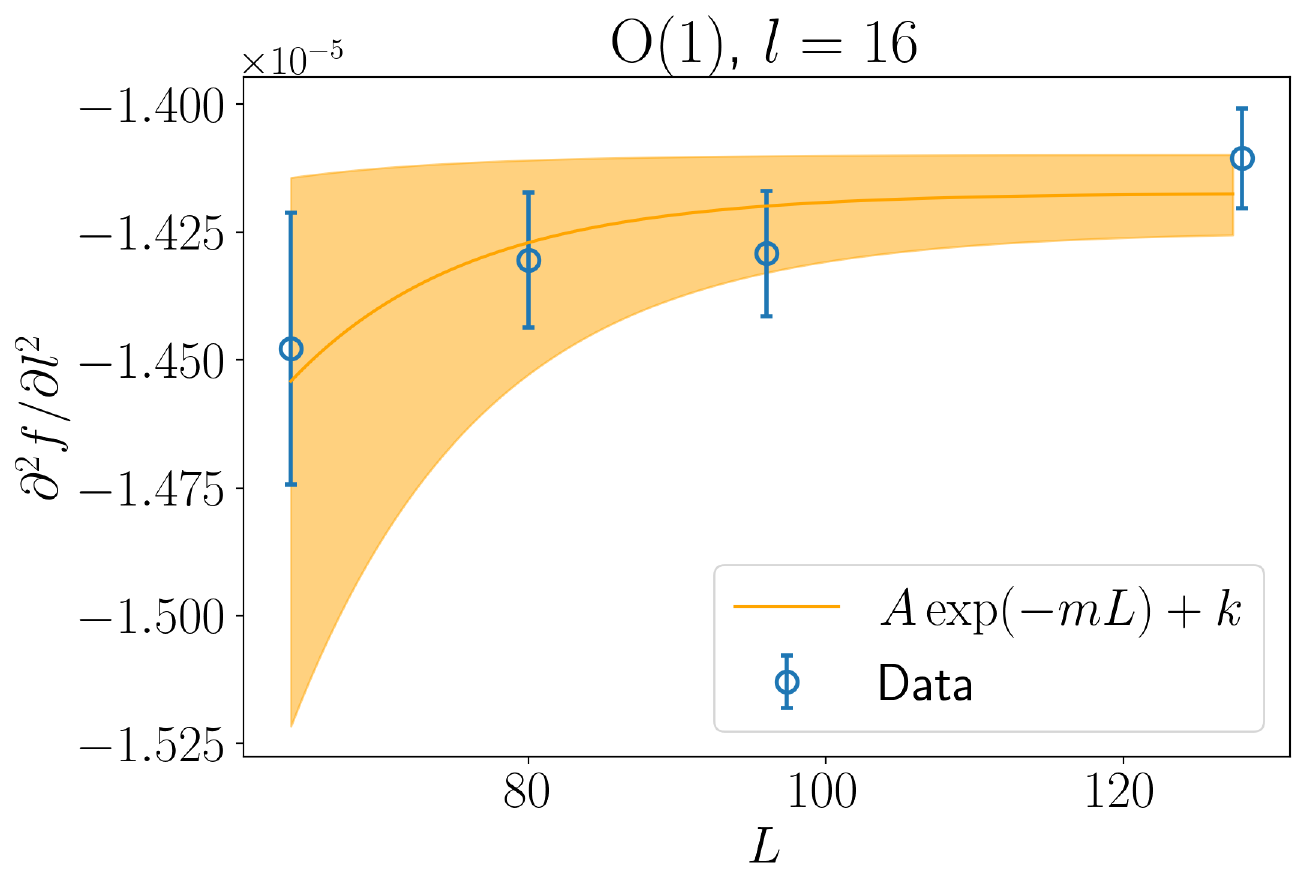}
    \end{subfigure}
    \caption{$L\rightarrow\infty$ extrapolation of $\partial^2\fex/\partial l^2$ results for $l=6,9,16$ for the $\Orth(1)$ model.}
    \label{fig:large_L_extrapolation}
\end{figure*}
\begin{table*}[ht]
    \centering
    \begin{tabular}{|c|c|c|c|c|c|}
    \hline
    & $\Orth(1)$ & $\Orth(2)$ & $\Orth(3)$ & $\Orth(4)$ & $\Orth(6)$ \\
    \hline\hline
    $\chi^2_{\mathrm{red}}[\#_{\mathrm{dof}}]$ & $1.65$ $[43]$ & $0.70$ $[30]$ & $1.16$ $[32]$ & $1.06$ $[23]$ & $1.48$ $[32]$\\
    \hline
    $m$ & $0.083(6)$ & $0.108(14)$ & $0.089(10)$ & $0.072(12)$ & $0.089(10)$\\
    \hline
    \end{tabular}
    \caption{Reduced $\chi^2$ and global fit parameters for different extrapolations in $L$ using eq.~\eqref{eq:large_L_fit_function}.}
    \label{tab:large_L_extrapolation}
\end{table*}
To extract the asymptotic behavior at large $L$ of $\partial^2 \fex/\partial l^2$ we performed an extrapolation using the fit function of eq.~\eqref{eq:large_L_fit_function}. For each value of $N$, we performed a combined fit of our data, with the coefficient $m$ in the exponent as a global parameter of the fit. Fig.~\ref{fig:large_L_extrapolation} presents representative fits for fixed $l$ at $N=1$, while in table~\ref{tab:large_L_extrapolation} the reduced chi squared and the best-fit values of $m$ are listed. For each dataset, the fitting window in $L$ was chosen to ensure compatibility, within uncertainties, of the, at least, two largest-$L$ estimates of $\partial^2 \fex/\partial l^2$.

It is worth noting that Vasyliev et al.~\cite{Vasilyev_2009} did not carry out an explicit extrapolation for $L\rightarrow\infty$. Rather, the quoted results are for fixed $l/L\equiv \rho =1/6$. As a benchmark of the slab-exchange algorithm, we computed the value of $\Delta$ for $N=1,2$ for $\rho=1/6$. With this analysis we found $\Delta/N = -0.1524(3)$ for $N=1$ and $\Delta/N = -0.1506(5)$ for $N=2$, in good agreement with the results of~\cite{Vasilyev_2009}. We stress that these are not the results quoted in the main text, as for those results we performed the $L\rightarrow\infty$ extrapolation discussed in this section.

\section{Fit in $l$}
\label{app:smalll}

\begin{figure*}[t]
    \centering
    \begin{subfigure}{.32\linewidth}
    \includegraphics[width=1\linewidth]{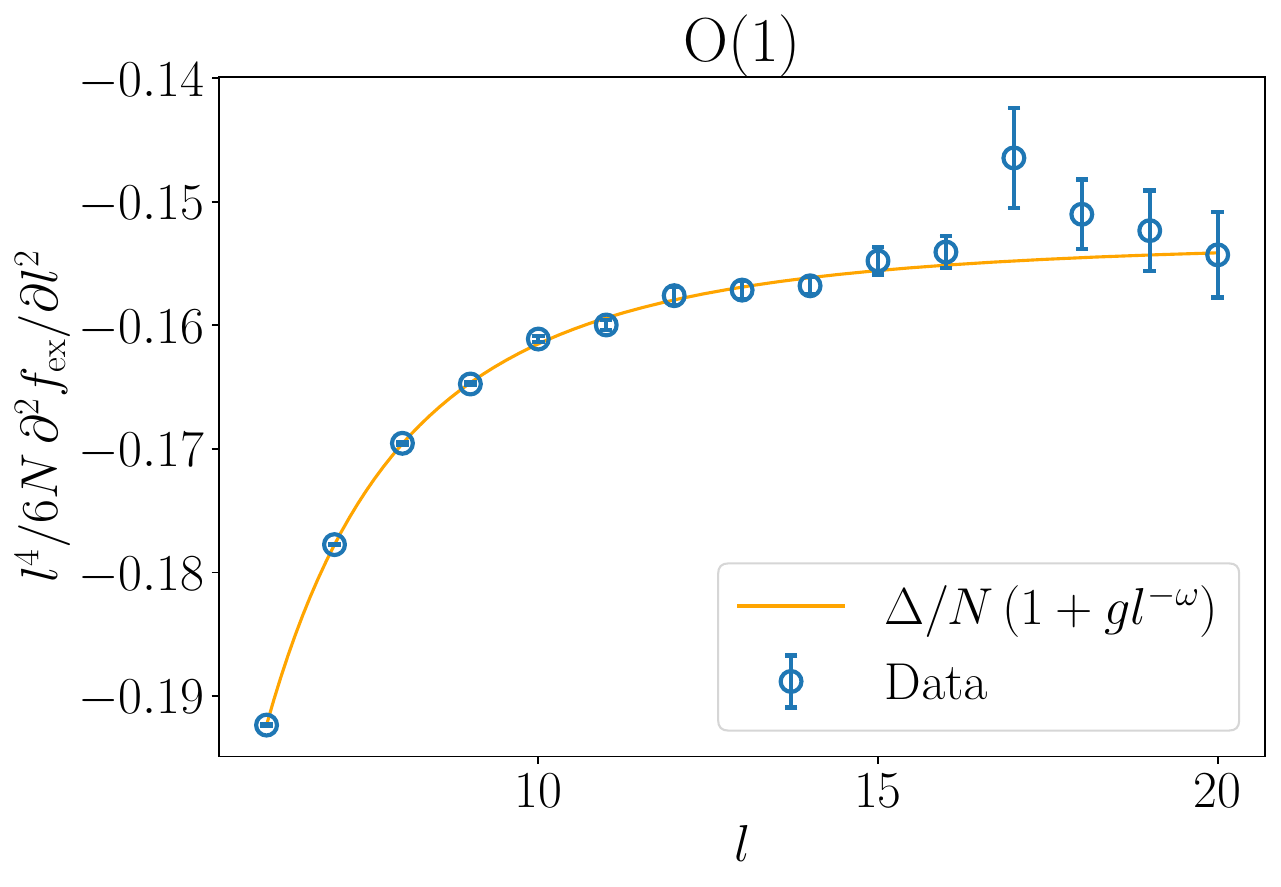}
    \end{subfigure}
    \begin{subfigure}{.32\linewidth}
    \includegraphics[width=1\linewidth]{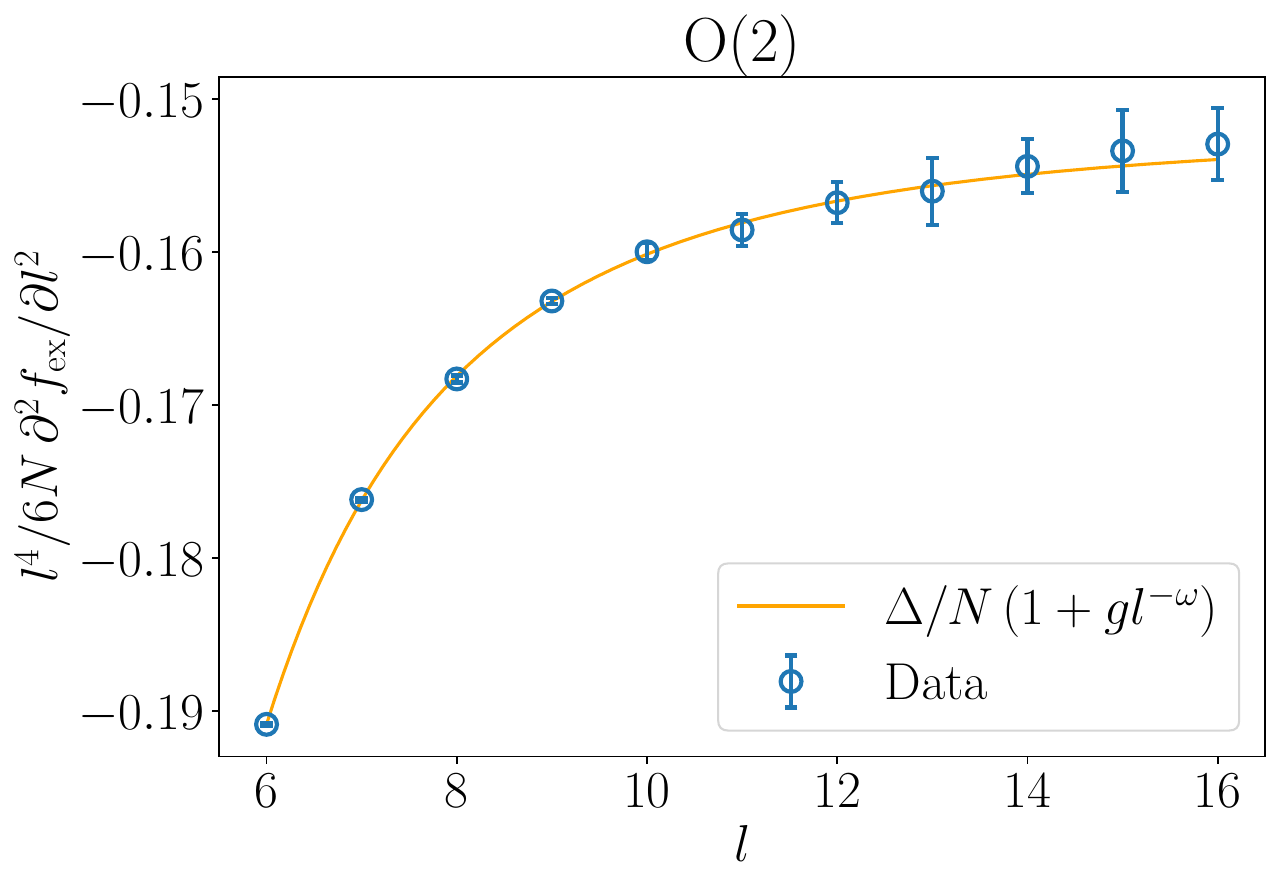}
    \end{subfigure}
    \begin{subfigure}{.32\linewidth}
    \includegraphics[width=1\linewidth]{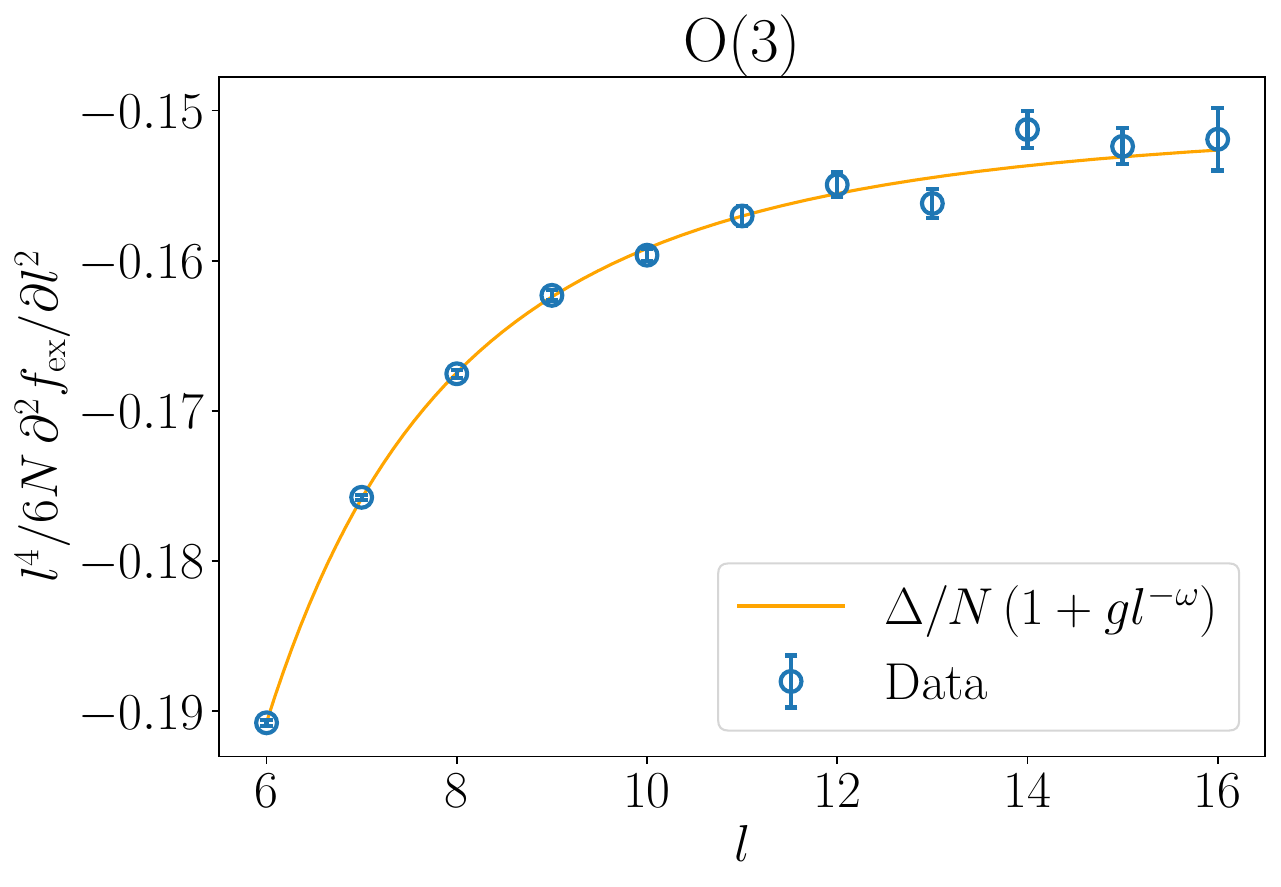}
    \end{subfigure}
    \begin{subfigure}{.32\linewidth}
    \includegraphics[width=1\linewidth]{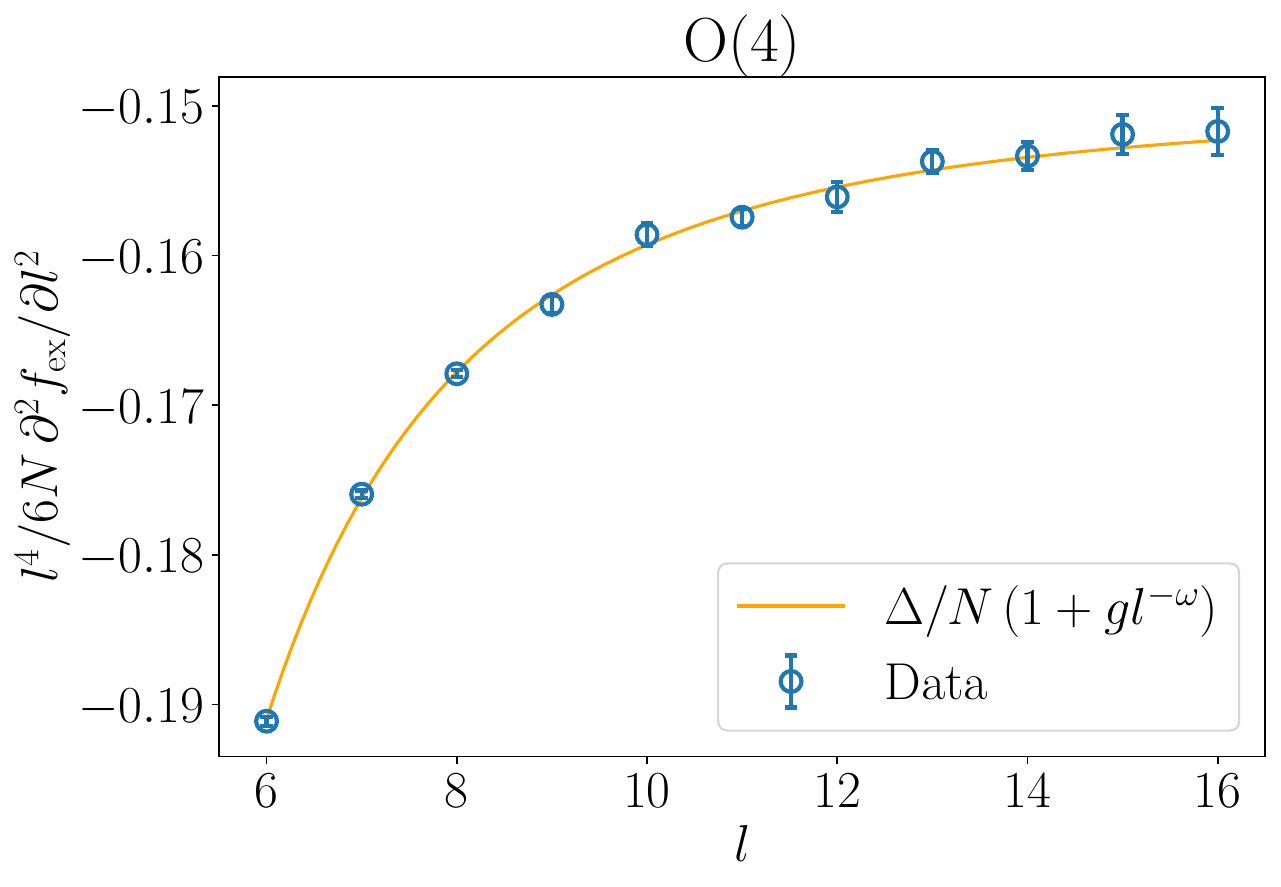}
    \end{subfigure}
    \begin{subfigure}{.32\linewidth}
    \includegraphics[width=1\linewidth]{Lz_fit_O6.pdf}
    \end{subfigure}
    \caption{Fit in $l$ to determine $\Delta/N$ for different $\Orth(N)$ models. Blue circles are all obtained with the previously discussed large-$L$ extrapolation.}
    \label{fig:fit_l}
\end{figure*}

Fitting the $l$-dependence of $\partial^2 \fex/\partial l^2$ is required to compute $\Delta/N$. In the scaling limit, the second derivative is expected to display a power law decay $6\Delta l^{-4}$. On the lattice one has to take into account scaling corrections arising from the finite extent of $l$ compared to the lattice spacing. Motivated by a finite size scaling analysis, in~\cite{Vasilyev_2009} the \textit{Ansatz} of eq.~\eqref{eq:fit_function_l} was introduced. The scaling correction $gl^{-\omega}$ has to be interpreted as the leading order of a series of progressively suppressed power-law corrections. As a consequence, small enough values of $l$ are not expected to be well approximated by the function of eq.~\eqref{eq:fit_function_l}, as higher-order corrections might be non-negligible.

In our analysis, we progressively excluded data for small $l$ until an acceptable value of the reduced $\chi^2$ is obtained. For all the values of $N$, we found that starting from $l=6$ consistently leads to a good fit. The results are reported in fig.~\ref{fig:fit_l} and in table~\ref{tab:fit_l}.

\begin{figure*}[t]
    \centering
    \includegraphics[width=0.4\linewidth]{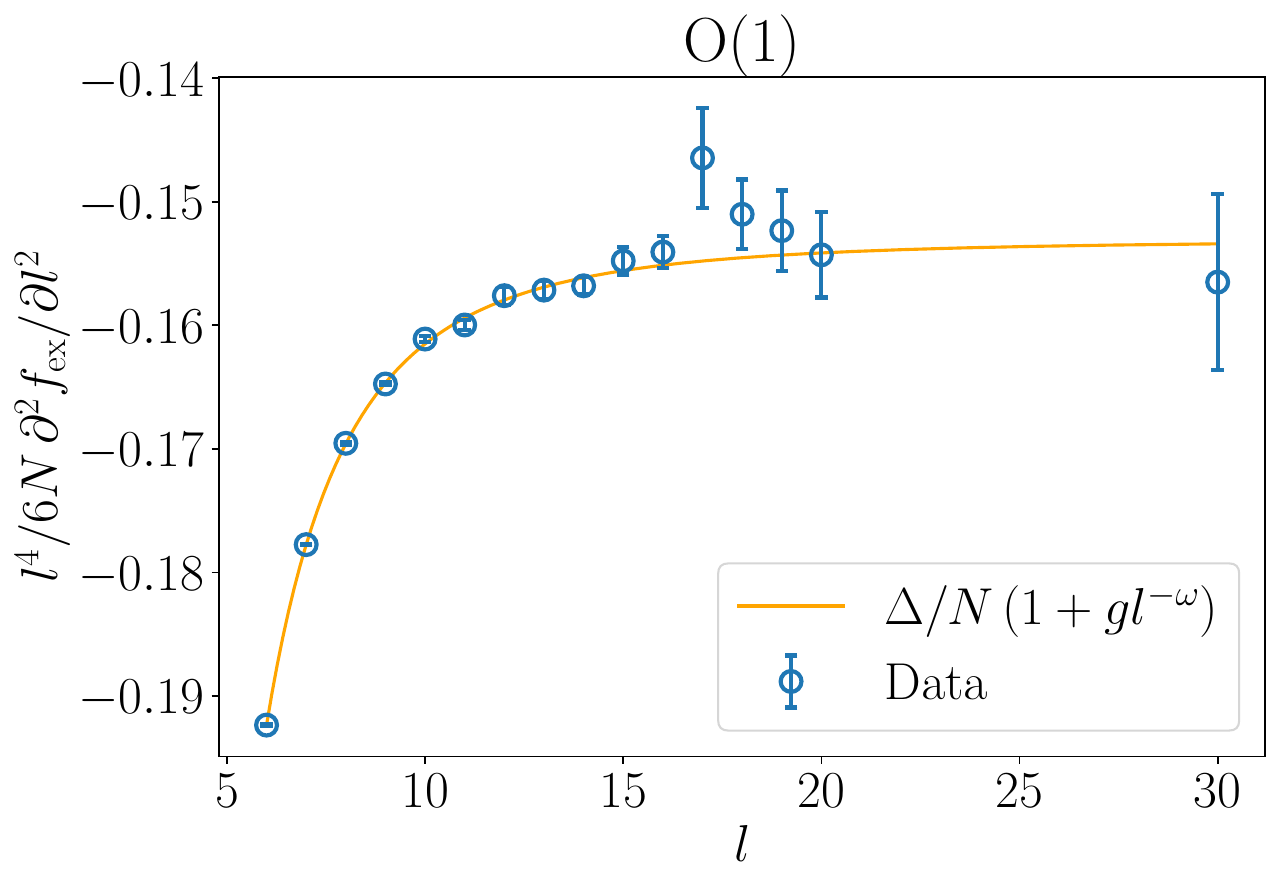}
    \caption{Fit in $l$ including a point at $l=30$.}
    \label{fig:fit_l_with_l30}
\end{figure*}

Finally, we checked potential systematic effects arising from the lattice sizes we used to perform the fit in $l$. In particular, for $\Orth(1)$ we limited ourselves to $l\leq 20$, while for $N>1$ we considered $l\leq 16$. Reference~\cite{Iliesiu:2018zlz} reports a Monte Carlo simulation of the Ising model on a $L=500$, $l=40$ lattice. In fig.~\ref{fig:fit_l_with_l30}, an additional point for $l=30$ is included. Even though we did not reach the same precision as for smaller values of $l$, the point clearly aligns with the others, and the result of the fit in $l$ is unaltered.


\bibliography{references}

\end{document}